## New Limits on Local Lorentz Invariance in Mercury and Cesium

S.K. Peck, D.K. Kim, D. Stein, D. Orbaker, A. Foss, M.T. Hummon and L.R. Hunter

*Physics Department, Amherst College, Amherst, Massachusetts 01002, USA*

We report new bounds on Local Lorentz Invariance (LLI) violation in Cs and Hg. The limits are obtained through the observation of the the spin- precession frequencies of $^{199}$Hg and $^{133}$Cs atoms in their ground states as a function of the orientation of an applied magnetic field with respect to the fixed stars. We measure the amplitudes of the dipole couplings to a preferred direction in the equatorial plane to be 19(11) nHz for Hg and 9(5) μHz for Cs. The upper bounds established here improve upon previous bounds by about a factor of four. The improvement is primarily due to mounting the apparatus on a rotating table. New bounds are established on several terms in the standard model extension including the first bounds on the spin-couplings of the neutron and proton to the *z* direction: $\tilde{b}_z^n < 7\times10^{-30}$ GeV and $\tilde{b}_z^p < 7\times10^{-29}$ GeV.

## I. INTRODUCTION

The postulates of Local Lorentz Invariance (LLI) and CPT invariance are cornerstones upon which nearly all of contemporary physics rests.[1] In 1995 some of the most stringent limits on possible violations of these symmetries were established by the Hg-Cs comagnetometer experiment.[2] The experiment compared the spin precession frequencies of Cs and Hg atoms as a function of the orientation of an applied magnetic field with respect to the fixed stars. Near the turn of the millennium, interest in such "clock comparison" experiments was renewed as the standard model extension (SME)[3] established a more consistent and coherent theory of LLI and CPT violation. The Hg-Cs comparison initially yielded the best limits on many of the SME parameters, including the equatorial dipole couplings $\tilde{b}_\perp$ for the electron, the neutron and the proton.[4,5] Subsequently, the limits on $\tilde{b}_\perp$ have been improved by other experiments.[6] The K-$^3$He experiment in Princeton[7] established limits on the neutron and proton of $\tilde{b}_\perp^n < 3.7\times10^{-33}$ GeV and $\tilde{b}_\perp^p < 6\times10^{-32}$ GeV while the Seattle torsion pendulum experiment[8] has established a bound on the electron of $\tilde{b}_\perp^e < 1.5\times10^{-31}$GeV. Two other labs have used $^{129}$Xe-$^3$He comagnetometers to establish limits on $\tilde{b}_\perp^n$ of $1.2\times10^{-31}$ GeV [9] and $3.7\times10^{-32}$ GeV.[10] An earlier torsion pendulum experiment placed a bound of $\tilde{b}_\perp^e < 3\times10^{-29}$GeV. [11]

Here we report the results of a second generation Hg-Cs LLI experiment. Several changes in the experiment, including the mounting of nearly the entire apparatus on a rotating table, have resulted in improved sensitivity. While refining the limits for LLI violation in $^{199}$Hg and $^{133}$Cs, the new result is unable to improve upon the present limits on $\tilde{b}_{\perp}$. However, the results do establish improved bounds on several other SME parameters. The rotating apparatus is sensitive to the previously unmeasured neutron and proton dipole couplings along the Earth's spin axis, $\tilde{b}_z^n$ and $\tilde{b}_z^p$. In addition, the most recent review of the field lists the original Hg-Cs comparison as yielding the best limits on the LLI violating SME parameters $\tilde{d}_X, \tilde{d}_Y, \tilde{g}_{DX}, \tilde{g}_{DY}$ for the proton ($< 10^{-25}$ GeV) and neutron ($<10^{-28}$ GeV).[12] Our experiment improves these bounds. The *d* coefficients are particularly interesting because unlike most of the other terms they are CPT-even; they break Lorentz symmetry without violating CPT. Recently, efforts have been made to disentangle various LLI violating terms of the SME by comparing clock experiments performed on heavy and light nuclei.[13] As Hg and Cs both have heavy nuclei, the present results may also prove useful in these efforts.

## II. THE HG-CS COMAGNETOMER

Conceptually, this experiment is very similar to the original Hg-Cs comparison and relies on the fact that the LLI violating terms that involve the electron only affect Cs, while Hg is primarily sensitive to the neutron. Our measurement is sensitive to Hamiltonians of the form $H = kh\hat{\boldsymbol{\sigma}} \cdot \hat{\boldsymbol{V}}$ where $h$ is Planck's constant, $\hat{\boldsymbol{\sigma}}$ is a unit vector in the direction of the atomic spin, and $\hat{\boldsymbol{V}}$ is a unit vector that might represent a preferred spatial direction or the particle's momentum with respect to a preferred frame, and $k$ is a constant that characterizes the strength of the LLI violating interaction. We measure the Larmor frequencies of Cs and Hg magnetometers as a function of relative orientation between an applied magnetic field, $\boldsymbol{B}$, and $\hat{\boldsymbol{V}}$. This frequency is given by $f = \gamma B + k\hat{B} \cdot \hat{V}$ where $\gamma$ is the atom's ground state gyromagnetic ratio ($\gamma_{Cs}$ = 350 kHz/G and $\gamma_{Hg}$ = 759 Hz/G). If the Hg and Cs atoms experience the same magnetic field, the difference in the "effective" magnetic field measured by the Cs and Hg magnetometers is given by

$$\Delta B = (k_{Cs} /\gamma_{Cs} - k_{Hg}/\gamma_{Hg})\, \hat{B} \cdot \hat{V}. \qquad (1)$$

Barring a fluke cancellation between the different particles, the comparison retains sensitivity to LLI violation of both species, while eliminating residual effects associated with magnetic field drifts in the apparatus.

Since Cs and Hg magnetometers have not been successfully demonstrated in the same cell, we compare the responses of three independent optically pumped magnetometers. The magnetometers are stacked at the center of our apparatus with two $^{199}$Hg magnetometers on the outside of the stack and a single Cs magnetometer in the center. The average magnetic field experienced by the two Hg magnetometers is an excellent approximation of the field experienced by the Cs magnetometer.

## A. The light absorption oscillators

The geometry of our optically-pumped light absorption oscillators is illustrated in Figure 1.[14] Circularly polarized laser light propogating along $\hat{z}$ (vertical in the lab) optically pumps the atoms in the presence of both static and oscillating magnetic fields, $\boldsymbol{B}$ and $\boldsymbol{B_{osc}}$. The static field is applied at an angle $\theta_B$ =63.8(1) degrees with respect to $\hat{z}$ and has a magnitude of ~ 5.5 $mG$. The horizontal component of $\boldsymbol{B}$ defines $\hat{x}$. The oscillating field is along $\hat{y}$ and modulates at two frequencies, $f_{Cs}$ and $f_{Hg}$, near the Larmor frequencies of Cs and Hg. Each magnetometer acts as a driven oscillator with a net atomic polarization, $\boldsymbol{P}$**,** precessing in a cone about $\boldsymbol{B}$ at frequency $f_{Cs}$ ($f_{Hg}$). The transmission of the circularly polarized laser light depends upon the projection of $\boldsymbol{P}$ along $\hat{z}$ and is therefore modulated in each cell at its corresponding frequency, $f_{Cs}$ ($f_{Hg}$). A change in the effective magnetic field (due to either a change in the real magnetic field or the LLI violating signal) results in a phase shift between the driving oscillatory magnetic field and the transmitted radiation. We lock the magnetic field to the Cs magnetometer and monitor the effective magnetic field experienced by the Hg magnetometers as a function of the orientation of $\boldsymbol{B}$ with respect to the fixed stars. This is essentially our LLI violating signal.

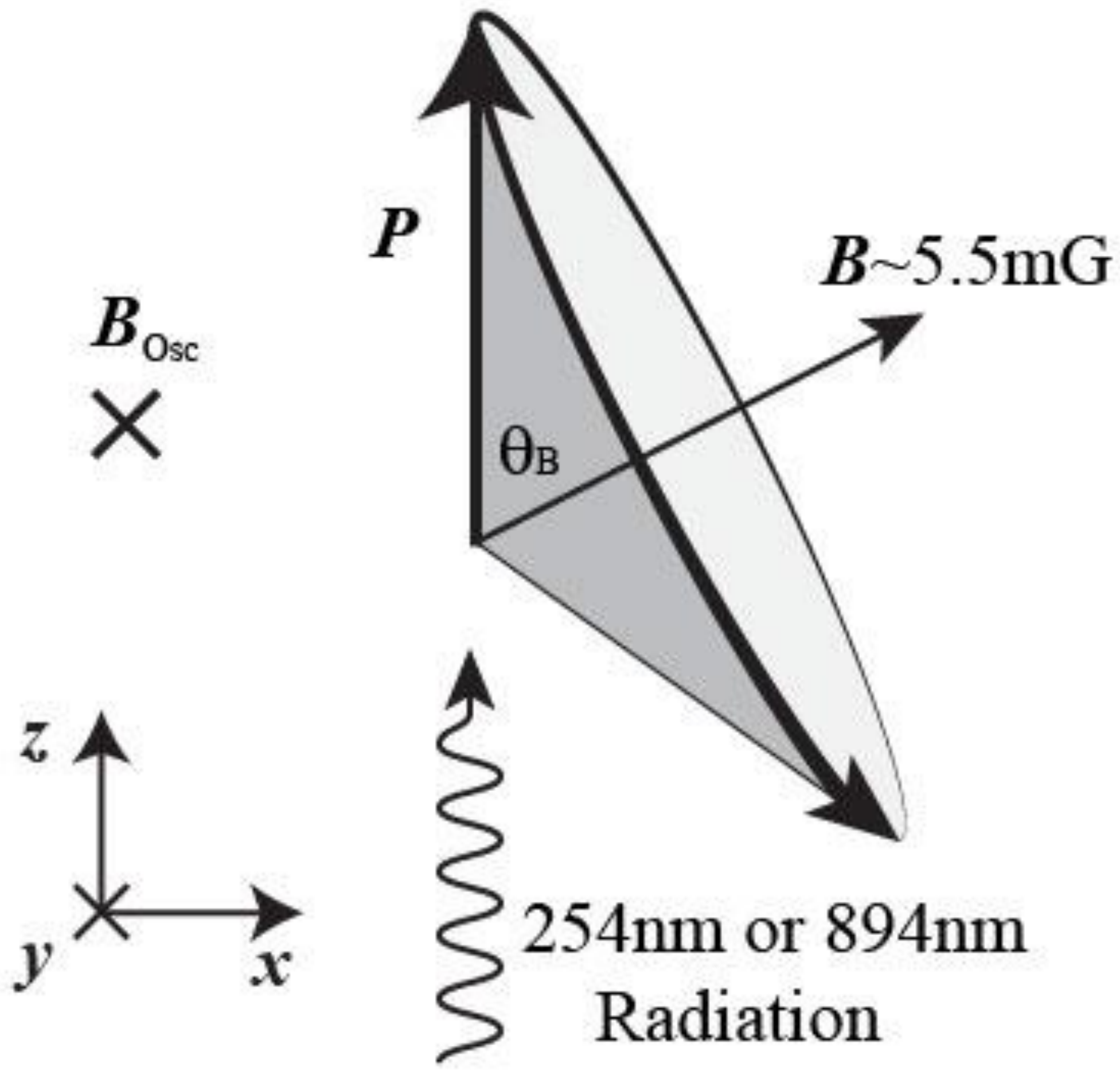

FIG. 1. Geometery of the light-absorption oscillators.

The atomic vapor cells are critical to achieving good magnetometer performance. The Cs cell is constructed from glass flats fused to a 1 cm section of 1.5" diameter Pyrex tubing. The interior cell walls are coated with high density paraffin to achieve spin relaxation rates of about 20 ms. There is no buffer gas in the cell so in a single spin relaxation time the atoms effectively sample and average the magnetic field over the entire volume of the vapor cell. The Hg vapor cells are similarly constructed but quartz tubing is used and 300 Torr of nitrogen is added to minimize wall collisions. A small amount of CO is added to quench the metastable 6 $^3P_0$. Due to their long transverse spin relaxation time of about 20 seconds, the Hg atoms also average the magnetic field over the entirety of their cells. This averaging of the magnetic field over the cells has virtually eliminated the coupling of magnetic gradients to the laser-beam position that plagued the original Hg-Cs comparison.

We use the 6 $^2S_{1/2}$ (F=4) – 6 $^2P_{1/2}$ (F=3) transition at 894 nm to drive the cesium magnetometer. About 1.3 mW of single-mode power is produced by a diode laser in an external cavity. The laser and all of its associated optics are mounted on the rotating table. The laser cavity is contained within a half inch thick aluminum box and surrounded by a layer of Sorbothane rubber and two one inch thick foam-insulation boxes. Combined with two stages of temperature control this configuration yields adequate vibration and temperature isolation. A small portion (~10$\mu$W) of the emerging laser beam is used for the experiment while the bulk is used to lock the laser frequency to a saturated absorption line.

We use the 6 $^1S_0$ (F = ½) – 6 $^3P_1$ (F = ½) transition for the optical pumping of $^{199}$Hg. In the original Hg – Cs comparison a $^{204}$Hg-discharge lamp was used to generate the required 254 nm ultraviolet (uv) radiation. We now generate the uv light using a Toptica Photonics frequency quadrupled diode laser system. An external cavity diode laser and tapered amplifier generates over half a watt of 1.016 micron radiation, which is frequency doubled using a $KNbO_3$ crystal in a ring cavity. Approximately 50mW of 508 nm radiation is transported via a single-mode polarization preserving optical fiber to the rotating table where a second doubling stage, using a BBO crystal, produces about 250 $\mu$W of 254 nm radiation. Most of this radiation is used in diagnostic and control systems while only a few $\mu$W are needed for the Hg magnetometers. A "noise eater" consisting of a Pockels cell sandwiched between linear polarizers reduces intensity noise to below one percent at frequencies below 10 kHz. We pass a portion of the 254 nm light through a Hg vapor reference cell (with no buffer gas) and monitor its transmission. The resonance lines in the experimental vapor cells are pressure shifted due to the buffer gas. A side lock in the reference cell conveniently yields light at line center in the experimental cells. These systems allow us to keep tight control of both the light intensity and frequency and we are thus able to stabilize the AC light shifts.

The three vapor cells are stacked along $\hat{x}$ at the center of three "inner" cylindrical magnetic shields (Fig. 2). A plexiglass cylinder fits snugly within the innermost shield. Three magnetic field coils wrapped on this cylinder are used to create the highly homogeneous fields required ($\boldsymbol{B}$ and $\boldsymbol{B_{osc}}$). Two additional coils, also supported by the cylinder, allow for the control of magnetic field gradients. Both the 894 nm and 254 nm laser beams pass vertically through the magnetic shields from bottom to top. The beams traverse their corresponding cells in the cell stack at the center of the shields. The transmitted radiation is guided out of the shields and detected by photodiodes that are fixed to the assembly. The 894 nm light transmitted through the Cs cell is guided by four optical fiber bundles arranged to detect four different quadrants of the beam. The light is measured on four photodiodes with carefully matched preamps. The intensity of these quadrants provides a sensitive means of detecting beam movement. A voltage proportional to the sum of the four photodiode currents provides the input signal for the cesium magnetometer's phase sensitive detection (Stanford Research SR850). For the Hg magnetometers, the transmitted 254 nm light is guided by a pair of aluminized Mylar tubes to two uv sensitive photodiodes. These signals are separately demodulated by another pair of SR850 detectors using the same reference signals that generate the oscillating magnetic fields used in the magnetometers. The lock-in reference signals for the Cs and Hg magnetometers, at 1.9198 kHz and 4.1665 Hz respectively, are both derived from a single 16 MHz clock. Using a single frequency source eliminates electronic sources of drift in the oscillator phases.

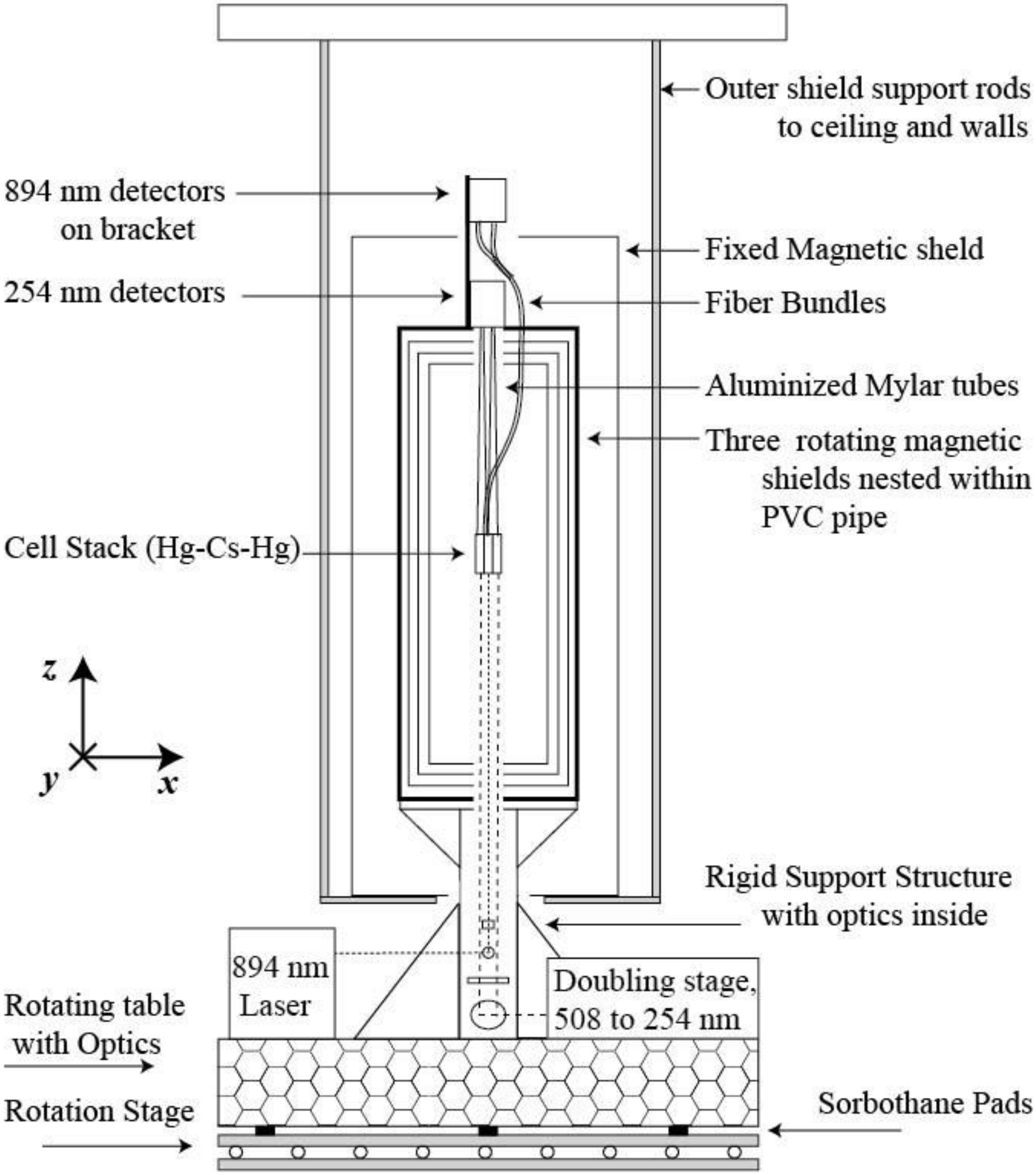

FIG. 2. The rotating comagnetometer apparatus.

## B. The rotation apparatus

The most important innovation introduced in the present experiment has been the mounting of nearly the entire apparatus on a rotating non-magnetic octagonal steel optics table (Fig. 2). The entire inner magnetic shield assembly is encased in a 43” long, thick walled PVC tube that provides structural stability. Both the inner shields and the PVC tube are supported by an aluminum pedestal. The pedestal rises from the optics table and consists chiefly of a 20” long ½” wall aluminum pipe with a 1” thick 43” diameter aluminum top. The pipe’s 4.5” outer diameter allows it to project up through the fixed magnetic shield end cap and support the 120 pounds of apparatus. The rigidity of the joints at the top and bottom of the tube are improved by gussets and cross struts to eliminate any tilting or twisting of the apparatus relative to the optics table. The octagonal optics table itself rests on a “lazy Susan” type rotation stage consisting of 2¼” diameter brass rollers sandwiched between two 1” thick aluminum plates. Twelve spokes projecting from a hub underneath the center of the optics table are used as axles for the rollers. The top plate of the rotation stage flexes as the supporting rollers change position. We decouple the optics table from this flexing by supporting it on three Sorbothane rubber pads.

To alter the direction of $\boldsymbol{B}$, the experiment is rotated between two table positions 180º apart. A reversible electric motor mounted over 7 meters from the experiment coils up a rope which is wrapped around and attached to the optics table. The table is thus pulled much like an old belt-drive turntable. Posts on the top of the lazy Susan contact a heavy plastic board fixed to the lab floor to create a positive stop with excellent reproducibility. The interior magnetic field assembly is parallel to local gravity to better than 1 arc minute for both positions of the table.

Molypermalloy magnetic shields like those used here are known to be much more effective at shielding static rather than dynamic magnetic fields. As such, when our apparatus is rotated, the changing direction of the ambient field with respect to the inner shields results in sub-optimal shielding. To address this problem, a stationary “outer” magnetic shield is suspended from the ceiling and surrounds the rotating “inner” shields. The outer shield reduces the ambient field (mostly the Earth’s magnetic field) experienced by the rotating shields to less than 1 *mG*. The residual ambient field within the inner most magnetic shield assembly is typically a few $\mu G$.

### C. The magnetometer signal on a rotating platform

The ability to rotate the apparatus is primarily responsible for our improved sensitivity. The direction of the magnetic field in the laboratory was fixed in the original Hg-Cs comparison and a full sidereal day was required in order to modulate the LLI signal. With this table assembly, rather than having to wait 24 hours for the rotation of the Earth, we can alter the direction of $\boldsymbol{B}$ every 9 minutes. This modulation of our expected LLI signal decouples it from other diurnal variations and substantially reduces the associated 1/f noise.

The component of the applied magnetic field that sweeps out the Earth's equatorial plane is sensitive to LLI violation in exactly the same manner as described in Berglund et al[2]. The component of the magnetic field aligned with the Earth's rotation axis is in a fixed direction in space. The atomic spin precession frequency about this axis is shifted by an amount proportional to the Earth's gyroscopic frequency, ($f_g$ = 1/(1 Siderial day) = 11.61 $\mu$Hz). This shift has little effect on the magnetic lock because the Cs magnetometer operates at nearly 2 kHz. However, it has a profound effect on the Hg magnetometers because of their lower precession frequency (about 4 Hz). All of our data are collected in two fixed positions labeled "signal" and "reference" (Fig. 3). In the "signal" position the horizontal magnetic field component points South in the lab while in the "reference" position, the horizontal component points North. The table rotation axis and the laser beams' propagation direction are vertical in our lab in Amherst, MA which has latitude $\theta_L$ equal to 42.370 degrees. To construct a data point we form the difference between the effective magnetic fields observed in the "signal" and "reference" positions. The rotation of the table leaves the vertical components of the field fixed and hence this difference is only sensitive to the change in the horizontal magnetic field which is proportional to $2B\sin(\theta_B)$. The sensitivity to the LLI signal is proportional to the projection of this horizontal magnetic field difference onto the equatorial direction (see equation 1), while the gyroscopic shift is proportional to its projection along the polar axis. Combining these effects we find that the LLI signal is proportional to $2\sin(\theta_B)\sin(\theta_L)$ while the predicted sensitivity of our apparatus to the Earth's gyroscopic frequency is $\nu_{pred} = C\,(1+ \gamma_{Hg} / \gamma_{Cs})\,(2f_g\sin(\theta_B)\cos(\theta_L) =$ 0.998*1.002 *2(11.61 μHz)sin(63.8)cos(42.37) = 15.39(2) μHz). The first term ($C$) in this expression accounts for a misalignment between the apparatus and true North, while the second term accounts for the small effect that the Earth's gyroscopic motion has on the Cs oscillator. The uncertainty in $\nu_{pred}$ is a result of the uncertainty in $\theta_B$. The frequency shift observed between the two table positions hence consists of the LLI signal that modulates with a period equal to one sidereal day superimposed upon the constant gyroscopic shift.

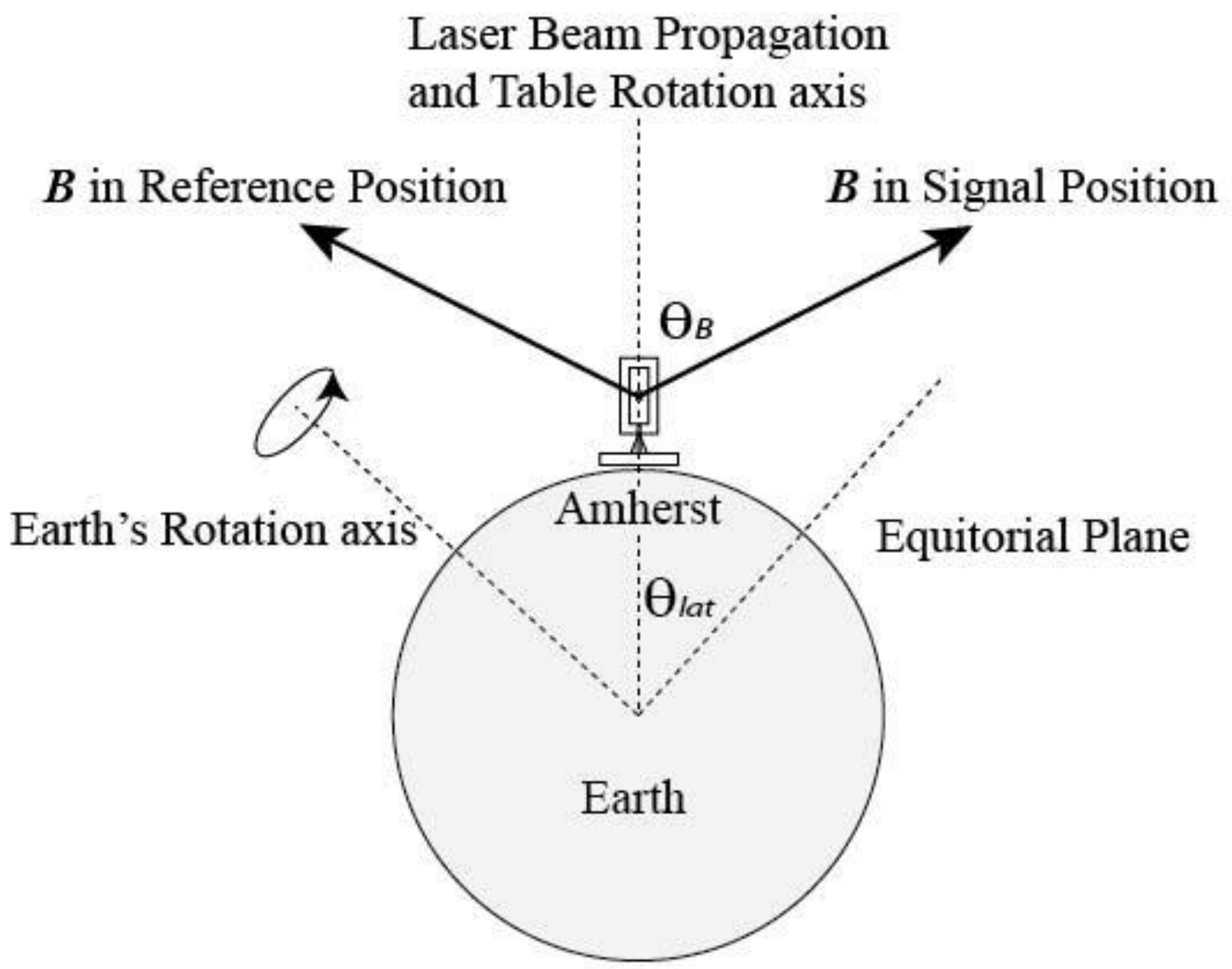

FIG. 3.  Both positions of the table have sensitivity to the LLI signal and to the Earth's rotation.

**D.  Optimization of sensitivity**

Several factors come into play in optimizing $\theta_B$.  The raw sensitivity of an ideal magnetometer in our geometry is proportional to $\sin(2\theta_B) = 2\sin(\theta_B)\cos(\theta_B)$ and optimizes at $\theta_B = 45$ degrees.  This is the angle that was used in Berglund et al[2].  If we  account for the sensitivity to the LLI effect, we have a signal that is proportional to $4\sin^2(\theta_B)\cos(\theta_B)\sin(\theta_L)$ which optimizes at $\theta_B = 65$ degrees.    Empirically, we found that our LLI sensitivity was reasonably well optimized for $\theta_B = 63.8(1)$ degrees, an angle that was conveniently produced by passing the same current through both the $\hat{z}$ (longitudinal) and $\hat{x}$ (transverse) magnetic coils.  All of the data presented here was collected with the applied magnetic field at this angle.  We note that should one have the ability to also adjust the latitude, latitudes nearer the poles benefit from both larger LLI sensitivity and reduced offsets from the Earth's rotation.

## III. DATA ACQUISITION

In order to extract the LLI signal, data must be accumulated for several days at a time. These long running times are made possible by a high degree of automation in the data acquisition process. Labview software, a PC and IEEE bus are used to control the experiment and record the data. In addition to the amplitude and phase of the oscillators, many other diagnostics signals are recorded including: room temperature, humidity, atmospheric pressure, ambient magnetic field components, laser intensity and frequency information and various control voltages. Data collection begins with adjusting the magnetic field and its gradient within the shields.

### A Calibration

To calibrate the experimental sensitivity it is necessary to determine the constant of proportion between the observed phase shift and the change in the precession frequency. To accomplish this, the current in our magnetic field coils is varied to produce a field change of 0.51 μG and the phase shifts of the driven oscillators are measured. The phase response to such small changes in current is highly linear.

### B. LLI data collection

Following calibration, the collection of LLI data begins. Data points are constructed from sets of four measurements. The phase of each magnetometer is averaged over a period of 124 seconds. The table is then rotated (this takes about 60 seconds) and after a wait period of 215 seconds, during which the Hg orientation catches up to the new direction of the magnetic field, the phase is averaged for two more 124 second intervals. The table is then rotated back to the initial position, and after a pause the phase is averaged again. This 18 minute sequence of four phase measurements constitutes a single data point and removes short-term linear drifts from the data. After collecting thirty data points, the entire cycle begins again with the resetting of the homogeneous and gradient fields and recalibration.

## IV. RESULTS

### A. Extraction of the LLI signal and the removal of magnetometer drift

In Fig. 4, we display data from a single data run as a function of time. In both table positions ("signal" and "reference"), the Hg magnetometer frequency (averaged over the two Hg cells) exhibits drift with time. However, when we construct a data point by taking the difference between these two frequencies (see section III. B), the drift is removed and the temporal variance in the frequency is substantially reduced. Additional noise reduction has been achieved by eliminating data points taken when there was excessive noise in laser intensity or frequency, or when the signal phase or amplitude drifted excessively during a single 124 second integration period. Such points can arise when acoustic noise disrupts the uv laser resulting in a jump in the phase and amplitude of the Hg oscillators. Such phase disruptions have a long recovery time, typically a few spin relaxation times.

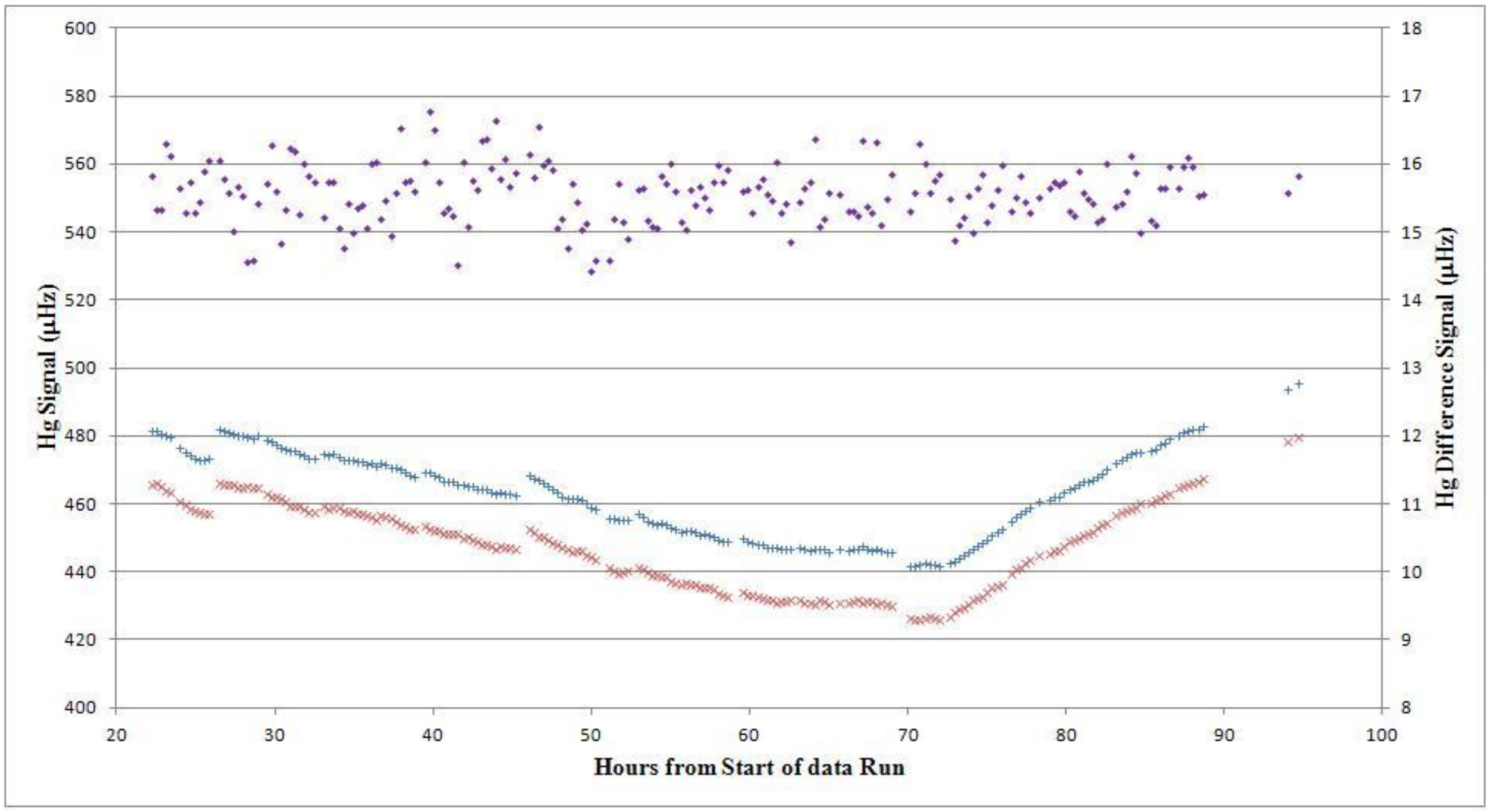

FIG. 4. The Hg signals for one data run (25-30 October, 2010) are shown for the two table positions: signal (pluses) and reference (exes). While there is a slow drift in the Hg signals, it does not appear in the difference between the table positions (diamonds) shown on the more sensitive secondary axis.

### B. Combining data from different data runs

We completed 12 data runs interspersed over a 13 month period (Table 1). Each data run was taken with a particular magnetic field orientation and typically consists of 5 days of data integration. To decouple our signal from possible diurnal variations, we alter the magnetic field orientation between runs. We choose one of four distinct combinations. The applied field is produced by a current running through two sets of magnetic field coils: one axial (along $\hat{z}$) and one along $\hat{x}$ (north-south). The reversal of the vertical field interchanges the "reference" and "signal" positions and changes the phase of the magnetic field direction with respect to the fixed stars by 180 degrees. The reversal of the horizontal field component interchanges the signal and reference positions but does not change the phase of the signal with respect to the fixed stars.

Table I. Dates and magnetic field configurations for the twelve data runs.

| | | Initial Magnetic Field Direction | | |
|---|---|---|---|---|
| Data Run | Dates | Axial | North-South | Data Points |
| 1 | Apr 1-8, 2010 | Down | North | 176 |
| 2 | Apr 12-15, 2010 | Down | South | 156 |
| 3 | June 15-24, 2010 | Up | South | 167 |
| 4 | Oct 4-15, 2010 | Up | North | 238 |
| 5 | Oct 20-25, 2010 | Up | South | 245 |
| 6 | Oct 25-30, 2010 | Down | South | 215 |
| 7 | Nov 1-6, 2010 | Down | North | 195 |
| 8 | Dec 6-10, 2010 | Down | North | 230 |
| 9 | Jan 5-9, 2011 | Down | North | 197 |
| 10 | Jan 10-15, 2011 | Up | North | 143 |
| 11 | Mar 7-11, 2011 | Up | North | 217 |
| 12 | Apr 4-23, 2011 | Up | South | 337 |

For each data run we average the frequency differences observed upon the rotation of table to extract the average frequency offset, $\nu_{offset}$ . These values for each data run are plotted in Fig. 5. There are clear non-statistical variations of these frequency shifts for different data runs. We have found two correlations between $\nu_{offset}$ and other table- dependent variables in the data set that we regard as significant. Both correlations are likely associated with small changes in laser intensity or frequency and corresponding changes in the ac light shifts[15] with the rotation of the table. One correlation is with the amplitude of the modulated Hg signals. This likely reflects an intensity dependent ac light shift in the Hg cells. The second correlation is between $\nu_{offset}$ and a signal that is a proxy for the Cs frequency, suggesting a frequency dependent ac light shift in Cs. The effect of removing these correlations is shown in Fig. 5. Due to an electronics failure it was only possible to correct run 12 for one of the two systematic effects. The uncertainty in run 12 was expanded to account for the likely range associated with the missing correlation. While statistical uncertainty in the corrections to the data increases the uncertainty for each individual data run, the corrected data is consistent from run to run and has an average value of $\nu_{offset}$=15.75(5) μHz.

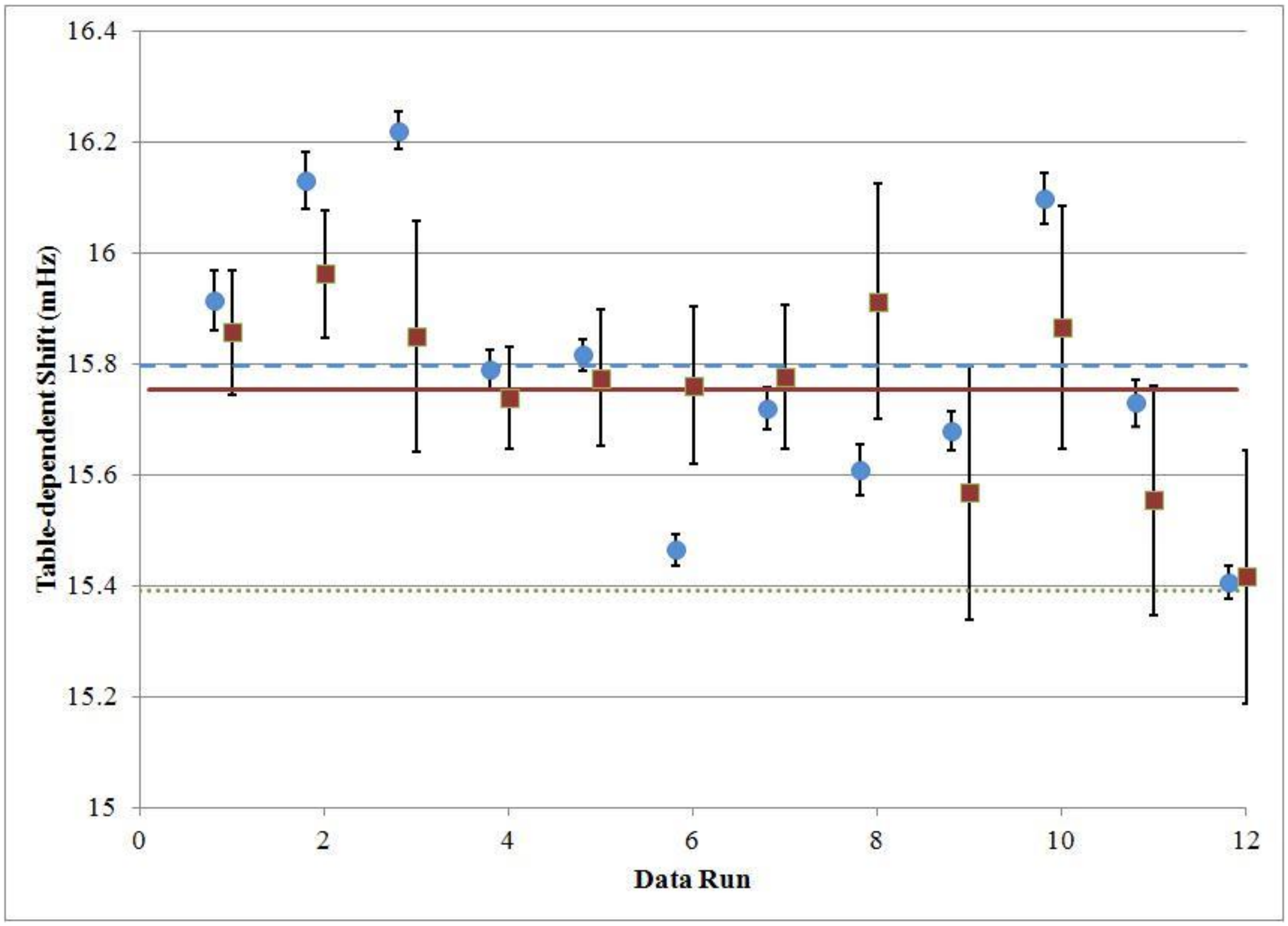

FIG. 5. Average table-dependent frequency shifts, $\nu_{offset,}$ for each of the 12 data runs. The dots correspond to the raw averages and the squares correspond to the data after correcting for the correlations discussed in the text. The dashed and solid lines correspond to the average of the raw and corrected data respectively. The dotted line shows $\nu_{pred}$, the offset expected from the Earth's gyroscopic effect.

The difference between $\nu_{offset}$ and $\nu_{pred}$ is $\Delta\nu_{gs}$ = 0.36 (7) μHz. We have investigated several possible sources of systematic error that may account for this difference. One possibility is that there is a quadrupole component to the magnetic field that is fixed in the laboratory frame and is imperfectly cancelled by the inner rotating shields. Such an effect could result in a differential shift of the Hg and Cs frequencies when the table is rotated. An imperfection in the stationary magnetic shield might be a source of such a field. We do not have a signal that permits us to distinguish this possibility from the observed signal. However, using the signals in the two mercury cells we are able to measure a table-dependent change in the gradient of the laboratory-fixed magnetic field. Averaged over our twelve data runs the table position-dependent frequency difference measured by the two Hg cells differs by 0.45 (16) μHz. For likely stationary magnetic sources, such as the welded seam of our outer shield, one would expect the frequency shift associated with the quadrupole field to be smaller than the dipole component by the ratio between the cell separations and the distance of the source. For our geometry this suppression is at least an order of magnitude, suggesting that this systematic is likely less than 0.05 μHz.

We suspect that the most likely source of $\Delta\nu_{gs}$ is associated with beam motion producing changes in the ac light shift. Despite extensive efforts to minimize beam motion when the table is rotated, some beam motion, about 60 μm, is still observed on the quadrant fiber bundles. This motion can result in a change in the light intensity (by etalon effects), beam direction (due to the curved cell surfaces and sub-optimal optical quality of our vapor cells), or of light polarization (due to position dependent cell birefringence). All three of these position-dependent effects can produce changes in the ac light shift that are synchronized with the table rotation. We have studied these effects and estimate that they may produce table dependent frequency shifts of about 0.3 μHz. Considering these results, we find that $\Delta\nu_{gs} = 0.36\ (7)\ (35)$ μHz, where the first uncertainty is statistical and the second is systematic. We interpret this result as an upper bound, $|\Delta\nu_{gs}| < 0.7$ μHz .

To extract the equatorial LLI signal we first subtract $\nu_{offset}$ from each data point within a run. To account for slow drifts in the frequency offset, a function quadratic in time is fit to each data set and subtracted from each data point. The resulting data (2,516 points) from our 12 runs are plotted in Fig. 6 as a function of the right ascension of the magnetic field direction.

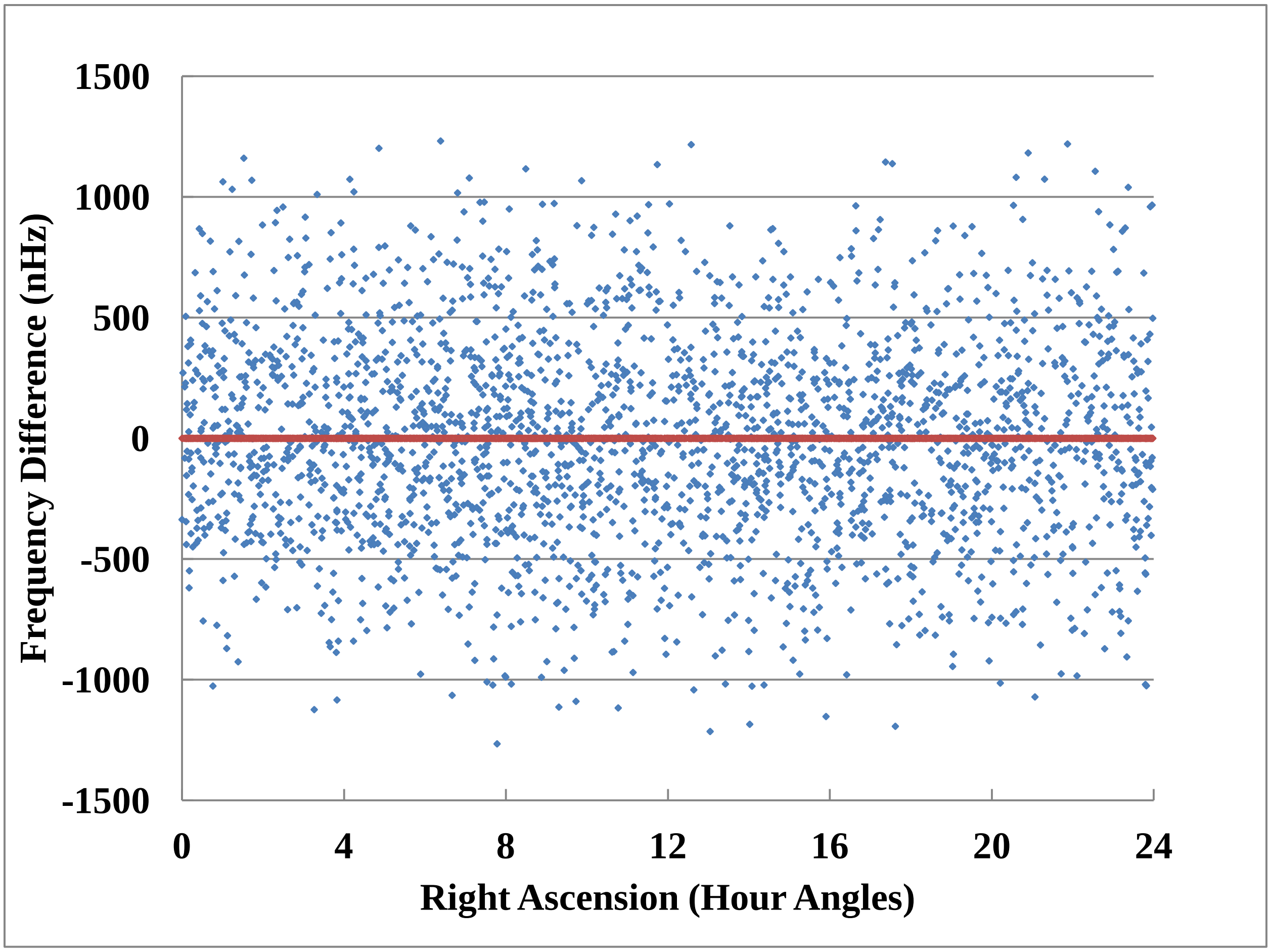

FIG. 6. The Hg frequency difference as a function of the direction of the magnetic field relative to the fixed stars (The Right Ascension in sidereal Hours). The best fit sinusoidal function is also shown (which is nearly indistinguishable from the *x*-axis in the figure.)

These data are fit to the function $A\cos(\omega t+\phi)$ where $\omega = 2\pi/(SiderealDay)$. The resulting fit parameters are $A$ = 19(11) nHz and $\phi$ = 5.4(6) radians where the quoted uncertainties are one sigma. A fit to the function $A\cos(\omega t)+B\sin(\omega t)$ yields the fit parameters $A$ = 12(12) nHz and B = 15(11) nHz. We find no significant correlations of these data with any of of the myriad of experimental and environmental parameters we have recorded. The systematic effects that we are aware of do not significantly contribute to this result at the present level of precision. The coefficient *A* above is a measurement of the possible coupling of the Hg atom to a preferred equatorial direction assuming that the Cs atom has no sensitivity. Conversely, if one assumes the Hg atom has no sensitivity, the above bound can be interpreted as a limit on the Cs coupling of 9(5) μHz .

### C. Implications for LLI violating SME coefficients

The null result for Hg can further be interpreted as an upper bound on the equatorial couplings of the neutron and proton[16] of $\tilde{b}_{\perp}^{n} < 3.7\times10^{-31}\,\text{GeV}$ and $\tilde{b}_{\perp}^{p} < 4\times10^{-30}$ GeV. These results are both about a factor of four smaller than the bounds that were established from the original Hg-Cs comparison. The bounds on the LLI violating SME parameters $\tilde{d}_X, \tilde{d}_Y, \tilde{g}_{DX}, \tilde{g}_{DY}$ for the proton and neutron are also improved by this same factor and are now about $3\times10^{-28}\,\text{GeV}$ and $3\times10^{-29}\,\text{GeV}$ respectively.

The strength of the LLI violating dipole couplings along the Earth's spin axis are denoted in the SME by the coefficients $\tilde{b}_z^{e}$, $\tilde{b}_z^{n}$, and $\tilde{b}_z^{p}$ for the electron, neutron and proton. These coefficients are notoriously difficult to measure since they are easily masked by the large frequency shift associated with the Earth's gyroscopic frequency. Of these three fundamental fermions that make up ordinary matter, only the electron's coupling has been previously bounded, $\tilde{b}_z^{e} < 4.4\times10^{-30}$ GeV.[8] Our limit on $\Delta\nu_{gs}$ establishes a 0.5 μHz bound on the frequency shift associated with this Hg spin coupling and implies that $\tilde{b}_z^{Hg} < 2\times10^{-30}$ GeV. We use the spin composition of Hg to infer limits on the proton and neutron couplings of $\tilde{b}_z^{n} < 7\times10^{-30}\,\text{GeV}$ and $\tilde{b}_z^{p} < 7\times10^{-29}\,\text{GeV}$.[16]

## V. FUTURE PROSPECTS

A new generation of this experiment could expect significant improvements in sensitivity. The dominant experimental noise and the largest systematic uncertainties in this experiment both appear to be associated with ac light shifts. The optical-pumping geometry we have employed is unfortunately first-order sensitive to these shifts. Alternative "pump-probe" geometries with no first order light shifts have demonstrated superior sensitivities.[17,18] By implementing a similar geometry we expect an improvement of about a factor of twenty in our experimental sensitivity . To refine our measurements of $\Delta\nu_{gs}$ we will in addition need to either monitor the quadrupole magnetic field component or develop a cell in which the Cs and Hg magnetometers can coexist. With these modifications and our present rotation apparatus it should be possible to control the various possible systematic uncertainties to the requisite precision. Such improved bounds on the equatorial signals in heavy atoms should make it possible to more precisely disentangle the various terms in the standard model extension. The additional systematic controls should also result in better measurement of $\tilde{b}_z^{n}$ and $\tilde{b}_z^{p}$ for the neutron and proton. Improvement beyond this factor of twenty may require aligning the table rotation axis with the Earth's spin axis in order to eliminate the large gyroscopic frequency shift. This alignment could be achieved either by moving the experiment to the South pole[7] or by sacrificing the alignment of the table rotation axis with local gravity.

## ACKNOWLEDGEMENTS

We wish to thank D. Krause Jr., R. Cann and N. Page for technical assistance and S.K. Lamoreaux for contributions to the early phases of this project. We thank M.V. Romalis and V.A. Kostelecky for useful suggestions. This work was supported by Amherst College and NSF grants PHY-0855465, PHY-0555715 and PHY-0244913.

---